\documentclass[10pt,a4paper]{article}
\def\bnt{b_{_{\mathrm NT}}}
\usepackage{jcappub}
\usepackage[ngerman,english]{babel}
\usepackage{bm}
\usepackage{dcolumn}
\usepackage{graphicx}
\usepackage{enumitem}
\usepackage{graphicx}
\usepackage{amssymb}
\usepackage{amsmath}
\usepackage{amsthm}
\usepackage{mathrsfs}
\usepackage{multirow}
\usepackage{bbding}
\usepackage{color}
\newif\ifAMStwofonts
\AMStwofontstrue 

\def\cl{{C_l}}

\def\al{{a_{lm}}}
\def\tal{{{\tilde a}_{lm}}}

\def\gsim{~\rlap{$>$}{\lower 1.0ex\hbox{$\sim$}}}

\def\simpropto{\lower.2ex\hbox{$\; \buildrel \propto \over \sim \;$}}
\def\ltsim{\lower.5ex\hbox{$\; \buildrel < \over \sim \;$}}
\def\gtsim{\lower.5ex\hbox{$\; \buildrel > \over \sim \;$}}
\def\ltsim{\lower.5ex\hbox{$\; \buildrel < \over \sim \;$}}
\def\gtsim{\lower.5ex\hbox{$\; \buildrel > \over \sim \;$}}

\def\kms{\mbox{km\,s$^{-1}$}}

\def\dd{\,{\rm d}}

\def\kms{\ {\rm km\,s^{-1}}}
\def\hmpc{\ {\rm h^{-1}Mpc}}

\def\hmmpc{\ {\rm h\,Mpc^{-1}}}

\def\dd{{\rm d}}

\def\pmb#1{\setbox0=\hbox{#1}%
\kern-.025em\copy0\kern-\wd0
\kern.05em\copy0\kern-\wd0
\kern-.025em\raise.0433em\box0}

\def\vr{\pmb{$r$}}

\def\hvn{\hat {\vr}}

\def\simlt{\lower.5ex\hbox{$\; \buildrel < \over \sim \;$}}
\def\simgt{\lower.5ex\hbox{$\; \buildrel > \over \sim \;$}}

\newcommand{\beq}{\begin{equation}}

\newcommand{\eeq}{\end{equation}}
\def\beqa{\begin{eqnarray}}
\def\eeqa{\end{eqnarray}}
\def\fixit#1{}
\def\hmpc{h^{-1}\,{\rm Mpc}}

\def\dd{{\rm d}}

\def\cN{{\cal N}}

\title{Revisiting the NVSS number count dipole}
\author[a]{Prabhakar Tiwari}
\author[a]{and Adi Nusser }
\affiliation[a]{Physics Department and the Asher Space Research Institute - Technion, Haifa 32000, Israel}
\emailAdd{ptiwari@physics.technion.ac.il}
\emailAdd{adi@physics.technion.ac.il}
\date{\today}
\abstract{We present a realistic modeling  of the dipole component of the  projected sky  distribution of NVSS 
radio galaxies. The modeling relies on mock catalogs generated within the context 
of $\Lambda$CDM cosmology, in the linear regime of structure formation.  
After removing the contribution from the solar motion, 
the mocks show that the remaining observed signal is mostly (70\%) due to structures 
within $z\lesssim0.1$. The amplitude of the model signal  depends on the bias 
factor $b$ of the NVSS mock galaxies. For sources with flux density, $ S> 15 \; \rm mJy$, the bias recipe  inferred 
from higher order moments is consistent with the observed dipole signal at $2.12\sigma$. 
Flux thresholds above $20 \; \rm mJy$ yield a disagreement close to the  $3\sigma $ level. 
A constant high bias, $b=3$ is  needed to mitigate  the tension to the  $\sim 2.3\sigma$ level. 
 }

\begin{document}
\maketitle
\section{Introduction}
\label{sec:int}

The standard cosmological paradigm  is based on the ``Einstein's Cosmological Principle'' \citep{Milne:1933CP,Milne:1935CP}, i.e. 
the approach to homogeneity and isotropy  on the largest observable scales.
Galaxies in the  NRAO VLA Sky Survey (NVSS) \citep{Condon:1998}
have been assumed to lie  at   redshift $z\sim 1$, corresponding to   a comoving distance 
of  $\sim 3.3\rm Gpc$ for  the $\Lambda$CDM cosmological background \citep{Planck:2015}. 
The galaxy clustering at these distances should contribute very little to sky maps of  
NVSS galaxy  and flux distributions  on  large angular scales.
The observed dipole component of these maps should be  almost entirely due to
Doppler and aberration effects \citep{Ellis:1984} caused by the    local solar motion as measured 
by the Cosmic  Microwave Background Radiation (CMBR) 
\citep{Conklin:1971,Henry:1971,Corey:1976,Smoot:1977,Kogut:1993,Hinshaw:2009}\footnote{The measured 
CMBR dipole magnitude is of order $\sim 10^{-3}$ 
\citep{Conklin:1971,Henry:1971,Corey:1976,Smoot:1977,Kogut:1993,Hinshaw:2009} 
and the corresponding speed is found to be 369$\pm$ 0.9 km s$^{-1}$ in the direction, 
$l=263.99^o\pm 0.14^o$, $b=48.26\pm 0.03^o$ in galactic
coordinates \citep{Kogut:1993,Hinshaw:2009}. In J2000 equatorial coordinates, 
the direction parameters are $RA=167.9^o$, $DEC=-6.93^o$.}. 
There have been several attempts to measure the dipole in radio catalogs 
\citep{Baleisis:1998,Blake:2002,Singal:2011,Gibelyou:2012,Rubart:2013,Tiwari:2014ni,Tiwari:2015np} 
and somehow the dipole measured from the  NVSS is puzzling 
with several independent analyses \citep{Blake:2002,Singal:2011,Gibelyou:2012,Rubart:2013, Tiwari:2014ni,Tiwari:2015np} 
yielding  values  which disagree
with the CMBR velocity dipole \citep{Singal:2011,Tiwari:2014ni,Tiwari:2015np} 
at $\sim 3 \sigma$ significance. 

The first extraction of the solar motion from the NVSS   dipole was done in \cite{Blake:2002}. 
The authors \cite{Blake:2002} reported a velocity which was roughly in the same direction as 
CMBR but  1.5 to 2 times larger in amplitude. However, nine years later, 
using the same data, a huge solar motion, ~1600$\pm$ 400 km s$^{-1}$,  
was reported \cite{Singal:2011} from both the projected number  density (per steradian) and brightness dipole measurements. 
This  significant ($>3\sigma$) disagreement with the observed solar motion, 
 immediately attracted the attention of the community resulting in  further  detailed analyses  of the  NVSS dipole
\cite{Gibelyou:2012};\cite{Rubart:2013}(\citep{Singal:2014});\cite{Tiwari:2014ni,Tiwari:2015np}.

The authors of \cite{Gibelyou:2012} derived  a dipole almost six times larger
than the CMBR velocity dipole,  which was  attributed
to observational biases.
Another analysis by \cite{Rubart:2013} yielded a dipole
with an amplitude roughly three times larger than the prediction from the CMBR value.
A more  recent analysis by \citep{Tiwari:2014ni} found a  $2.7 -3.2\sigma$ 
disagreement between the NVSS  number  dipole and  the CMBR for various flux density cuts. 
Conclusively, the present status of NVSS dipole is a disagreement with CMBR 
predicted velocity dipole at $2.7 \sigma$ or more. 

However,  the contribution of the   intrinsic variation 
of the projected number  density as a result of the NVSS galaxy clustering remains to be properly assessed.
The  intrinsic  dipole  mainly depends on the following key ingredients:
\begin{itemize} 
\item number density distribution $N(z)$ as a function of redshift. This dictates  the fractional 
contribution of the large scale structure from different redshifts.  Available constraints on $N(z)$ strongly 
indicate that it peaks at redshifts lower than $z\sim 1$. 
\item relation between the  clustering of radio galaxies versus the underlying mass as a function of redshift. 
This biasing relation can be parametrized in terms of the galaxy, $P_{g}(k)$,  and matter power spectrum, $P(k)$, as, 
$P_{g}(k,z)=b^2(z) P(k,z) $. The bias factor, $b$,  could be as large as $2-3$ and hence produce a non-negligible  boost
to the dipole signal, especially if $N(z)$ peaks at redshifts lower than $z\sim 1$.

\item constraints from the Local Universe.
Our cosmological neighborhood within  100 Mpc  is moving with $\sim$300 km s$^{-1}$ \citep{Adi:2011vl} and roughly 
aligned with the solar motion. This motion is caused by even farther structures, producing an alignment 
between the solar motion and intrinsic dipole. 
\end{itemize} 
We aim at a realistic assessment of the NVSS dipole within the context of 
the $\Lambda$CDM model taking into account the above ingredients.
In  Ref. \cite{Adi:2015nb} (hereafter NT15), the authors  approximate the observed 
redshift distribution, $N^{\rm obs}(z) $ from  the best available redshift catalogs Combined EIS-NVSS
Survey  of Radio Sources (CENSORS)\citep{Best:2003,Rigby:2011} 
and Hercules \cite{Waddington:2000,Waddington:2001}.  They extract the  biasing factor $b(z)$  
 by matching the $\Lambda$CDM predictions to the NVSS measured angular power 
spectrum. The redshift dependent biasing  and number distribution are also found to
be consistent with the observed radio luminosity functions \citep{Mauch:2007,Smolcic:2008,Rigby:2011}. 
The  results obtained in NT15 are based on   high order multipoles $l\geq4$, 
where the analysis is fairly insensitive to  complications related to the partial sky coverage. 
 In this paper we focus on the dipole, $l=1$, and perform the comparison via $\Lambda$CDM mock catalogs.

The outline for the paper is as following. In \S\ref{sc:data}, we briefly describe the  NVSS catalog. 
In \S\ref{sc:proc} we review the basic formalism for describing the dipole from a projected map of galaxies.
We provide our treatment for the number density $N(z)$ and galaxy biasing in \S\ref{sc:nzbz} \&  \S\ref{sc:bz}, 
respectively. \S\ref{sc:sim} describes the NVSS mock catalogs. We present our results in  \S\ref{sc:res} and 
conclude with a general discussion  in \S\ref{sc:con}.

\section{Data}
\label{sc:data}

The full NVSS catalog
contains $\sim$1.7 million sources observed at 1.4GHz with an  integrated flux density $S_{\rm 1.4GHz}>2.5$ mJy.
The full width at half maximum angular resolution is 45 arcsec and nearly all observations are
at uniform sensitivity. Most of the sources (90\%)  are unresolved. 
The catalog covers the sky north of declination $-40^\circ$ (J2000), which is almost
82$\%$ of the celestial sphere. However, the effective sky coverage  for 
our study here is $75\%$  since we impose the following two cuts: 1)  we mask out  sources at latitudes $|b|<5^\circ$
to avoid Galactic contamination and 2) we remove additional 22 sites around bright
local extended radio galaxies identified by \cite{Blake:2002}, to avoid overwhelming
correlations which may leak to the large scales of interest. 
These are the only cuts we impose on the data. 
 The authors of \cite{Tiwari:2014ni}  have  studied the robustness of  the measured NVSS dipole with respect to 
 additional  cuts.
They have demonstrated  that  masking out latitudes of $<10^\circ$,  around the Galactic and Super-Galactic planes changes the 
dipole  by less that  $10\%$.
The same authors  have also assessed the effect of  removing sources within 30 arcsec of
known nearby galaxies listed in \cite{Vaucouleurs:1991,Corwin:1994,Saunders:2000,
Jarrett:2003,Huchra:2012,Bilicki:2014}. This procedure also resulted  in small changes 
in the NVSS dipole.

As shown by Ref. \cite{Blake:2002}, the NVSS suffers from significant systematic mis-calibration across the sky 
 at flux densities $S \ltsim$ 15 mJy, which affects  lower multipoles including the dipole. Therefore, we 
restrict our analysis to sources brighter than 15 mJy and consistently give the results for S $>$ 15, 20, 30, 40 and 50 mJy. 
Furthermore, for completeness and comparison to other authors 
\citep{Blake:2002,Rubart:2013,Tiwari:2014ni}, we give results for $S>10$ mJy, 
for one set of mocks with most conservative parameters. We impose a maximum flux density 
cut $S<1000$ mJy to remove extra bright sources. 
\section{Basics}

\subsection{The dipole component}
\label{sc:proc}

Let $\cN(\hvn)$ be  the projected number density (per steradian) in the line-of-sight direction $\hvn$,  and 
$\bar \cN$  be the mean number density averaged  over the sky. 
Further, let the density contrast   $\Delta (\hvn)\equiv \cN (\hvn)/\bar \cN- 1$.
Given the measured  angular positions of galaxies, we employ the HEALPix \citep{Gorski:2005}  scheme  to obtain
$\cN(\hvn)$ in equal area pixels.
We   expand   $\Delta (\hvn)$ in 
spherical harmonics, $Y_{lm}(\hvn)$, as 
\begin{equation}
\label{eq:delta}
\Delta (\hvn)= \sum_{l=1}^{\infty} \sum_{m=-l}^{+l} a_{lm}Y_{lm}(\hvn)\; .
\end{equation}
For a full sky coverage,  the  coefficients, $a_{lm}$ can easily be calculated by inverting 
equation (\ref{eq:delta}), as 
\begin{equation}
\label{eq:alm}
\al= \int_{4\pi} d \Omega  \Delta(\hvn) Y_{lm}(\hvn).
\end{equation}
The angular power, $\cl$, for multipole $l$ is, 
\begin{equation}
\label{eq:ens}
\cl=\langle |\al|^2\rangle \; .
\end{equation}
For a partial sky coverage, 
we resort to an approximate scheme   \citep{Peeb80}
to compute $\cl$, 
\begin{equation}
\label{eq:cobs}
C^{\rm obs}_l=\frac{\langle  |a^{\prime}_{lm}|^2\rangle}{J_{lm}} -\frac{1}{\bar \cN}
\end{equation}
where  $a^{\prime}_{lm}$ is obtained from  equation (\ref{eq:alm}) but with  angular integration over survey region only. Further, 
$J_{lm}=\int_{_{\rm survey}}|Y_{lm}|^2 \dd \Omega$ is the approximated correction suggested in 
reference \citep{Peeb80}. The term $\frac{1}{\bar \cN}$ describes the contribution of the 
Poissonian shot-noise.  
The $C^{\rm obs}_l$ is a measure of anisotropy at angular scale  $\sim \pi/l$. 

\subsubsection{The solar motion contribution to the  dipole}
\label{sc:vecd}
A moving  observer sees a brighter  sky 
in the  forward direction due to Doppler boosting and aberration effects  \cite{Ellis:1984}. 
 The flux density (energy per area per frequency) of sources in NVSS data follows 
a power-law  dependence on frequency $\nu$, $S \propto\nu^{-\alpha}$,
with $\alpha\approx 0.75$ \citep{Ellis:1984}.
Let $ v$ be speed of the observer relative to a source emitting radiation  intrinsic  frequency 
$\nu_{\rm rest}$. To lowest order in $v/c$, the observer sees the radiation at a frequency
$\nu_{\rm obs} =\nu_{\rm rest} \delta$, where
\begin{equation}
\delta \approx 1+(v/c) \cos \theta \  , 
\label{eq:deltad}
\end{equation}
and  $\theta $ is the angle between $v$ and the line of sight to the source.
Therefore, the observed flux density is  \citep{Ellis:1984},
\begin{equation}
S_{\rm obs}=S_{\rm rest} \delta^{1+\alpha}
\label{eq:Sobs}
\end{equation}
at a fixed frequency in observer's frame. The extragalactic radio source population count above some 
flux density cut, follows a power law $n(>S)\propto S^{-x}$ \citep{Ellis:1984}, where 
 $x\approx 1$  \citep{Tiwari:2014ni}.  Therefore,  the number 
count above some $S_{\rm obs}$ will change  by a factor $\delta^{(1+\alpha)x}$ due to Doppler effect. 
In addition, the  aberration of light  changes the solid angle as,
\begin{equation}
 d\Omega_{\rm obs}=d\Omega_{\rm rest} \delta^{-2}.
\label{eq:Sang}
\end{equation}
Using these equations, the dipole in the projected number  density 
due to the local motion, at leading order in $v/c$, is  \citep{Ellis:1984,Tiwari:2014ni}, 
\beq
\label{eq:D_n}
\vec D = [2+x(1+\alpha)](\vec v/c)\; .
\eeq

The dipole term of the power spectrum, $C_1$,  corresponds to  
to  a dipole amplitude, $D$ ($|\vec D|$) as in  equation (\ref{eq:D_n}),  given by \citep{Gibelyou:2012},
\beq
\label{eq:c2d}
C_1 = \frac{4\pi}{9} D^2.
\eeq

\subsubsection{Contribution from large scale structure}

In linear theory for gravitational instability, the theoretical  angular power spectrum ${\tilde C}_l$ is given in 
terms of the 3D mass density power spectrum $P(k)$ at $z=0$  as  \cite{Adi:2015nb}
\begin{eqnarray}
\label{eq:clphi}
{\tilde C}_l&=&<|\tal |^2> \nonumber\\
          &=&   \frac{2}{\pi }\int dk k^2 P(k) W^2(k) \; ,
\end{eqnarray}
where $W(k)=\int_{0}^{\infty} D(z) b(z) p(r) d r  j_l(kr)$ is the window function in $k$-space 
and $P(k)$ is the $\Lambda$CDM mass density power spectrum. Further, $D(z)$ is the linear growth 
factor normalized to unity at $z=0$ and $p(r)\dd r \propto N(z) \dd z$ is the 
probability for finding a galaxy in the range $r-(r+\dd r)$ and $r$ is the comoving distance corresponding to 
redshift $z$.

\subsection{The redshift distribution} 
\label{sc:nzbz}
\begin{figure}
\includegraphics[width=1.0\textwidth]{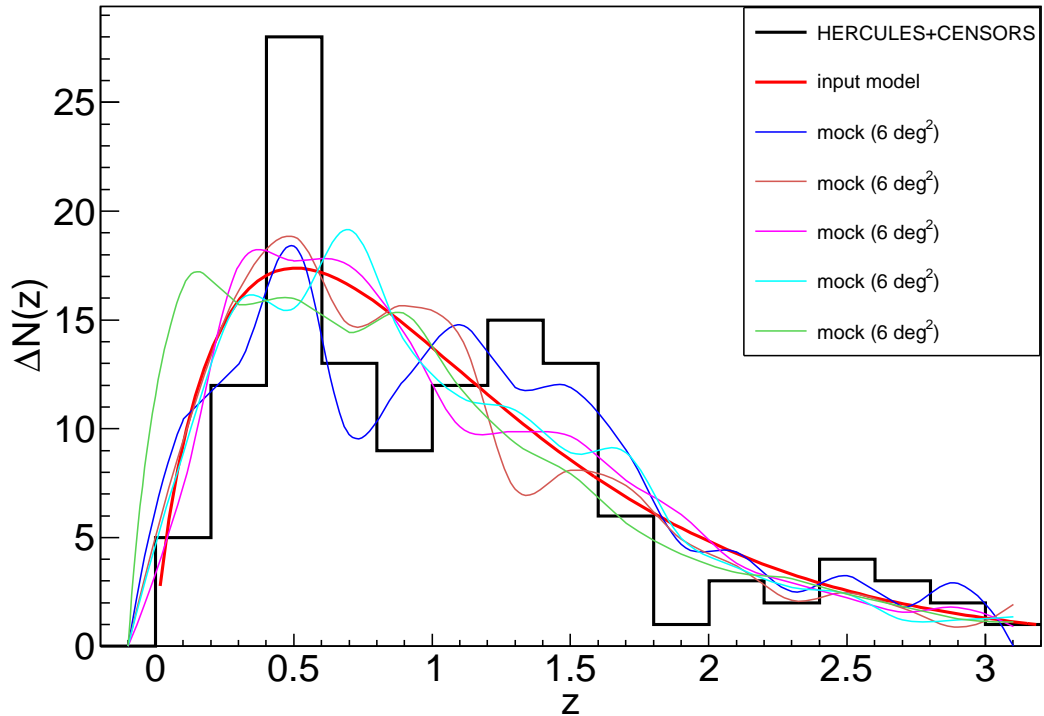}%
\caption{The histogram is the observed number of sources from combined CENSORS+Hercules 
data sets with flux density $S>10$ mJy and per redshift bin $\Delta z=0.2$. 
Best fit model (input model) follows  equation 
\eqref{eq:Nmodel} with parameters $a_1=0.74$, $a_2=0.71$ and $a_3=1.06$. The 
number density from mock catalogs for same area as observation is also shown. 
Note that the mocks include the $\Lambda$CDM density fluctuation, biasing $b(z)$ 
and the Poisson shot-noise.}
 \label{fig:NzSim}
\end{figure}

We rely on the results of  NT15 for biasing and  the redshift distribution for radio galaxies. 
These authors modeled the  redshift distribution of NVSS galaxies on the basis of  the 
observed distribution $N^{\rm obs}(z)$ obtained from 
CENSORS and   Hercules surveys. The CENSORS catalog was developed 
to study the evolution of the steep-spectrum radio luminosity function and the
survey is presumably complete for  flux densities $S>7.2$mJy at 1.4 GHz, containing 
135 sources. Indeed, the catalog covers only $6\rm deg^2$ of the ESO Imaging Survey 
Patch D (EISD). The redshift measurements of the sample are 73\% spectroscopically complete 
and for the remaining sources the redshift is  estimated by Kz or 
Iz magnitude-redshift relation. 
The Hercules redshift survey of 64 sources within 1.2 $\rm deg^2$ 
is complete for $S>2\rm mJy$ at 1.4 GHz.
We combine CENSORS+Hercules data to model NVSS source redshift distribution. 
We find 133 sources above 10 mJy flux density $S$. For higher flux density cuts we use the same as 
the number of sources reduces significantly and  effectively there  is no change in fit.  
We add that the CENSORS+Hercules combined data sets have been used to study 
the redshift dependent properties\citep{Best:2003,Rigby:2011,Adi:2015nb}.
NT15 use the parametric model 
\begin{equation}
\label{eq:Nmodel}
N^{\rm model}\propto z^{a_1}\exp\left[ -\left(\frac{z}{a_2}\right)^{a_3} \right] \; ,
\end{equation}
where the parameters  $a_1$, $a_2$ and $a_3$ have been  fixed by fitting the CENSORS+Hercules 
redshift distribution as well as the NVSS angular power spectrum for $l>4$.
The  best fit values  were found to be $a_1=0.74\pm 0.57$, $a_2=0.71\pm0.79$ 
and $a_3=1.06\pm0.53$.
A histogram of the  observed distribution, $N^{\rm obs}(z)$, is plotted  in figure  \ref{fig:NzSim} together with 
the  model fit in equation (\ref{eq:Nmodel}) for the best fit values of $a_i$. Further we have plotted
5 curves representing $N(z)$, for  $a_i$ randomly drawn from the full probability distribution  function computed in NT15.

\subsection{Galaxy biasing } 
\label{sc:bz}
\begin{figure}[!htb] 
\includegraphics[width=1.0\textwidth]{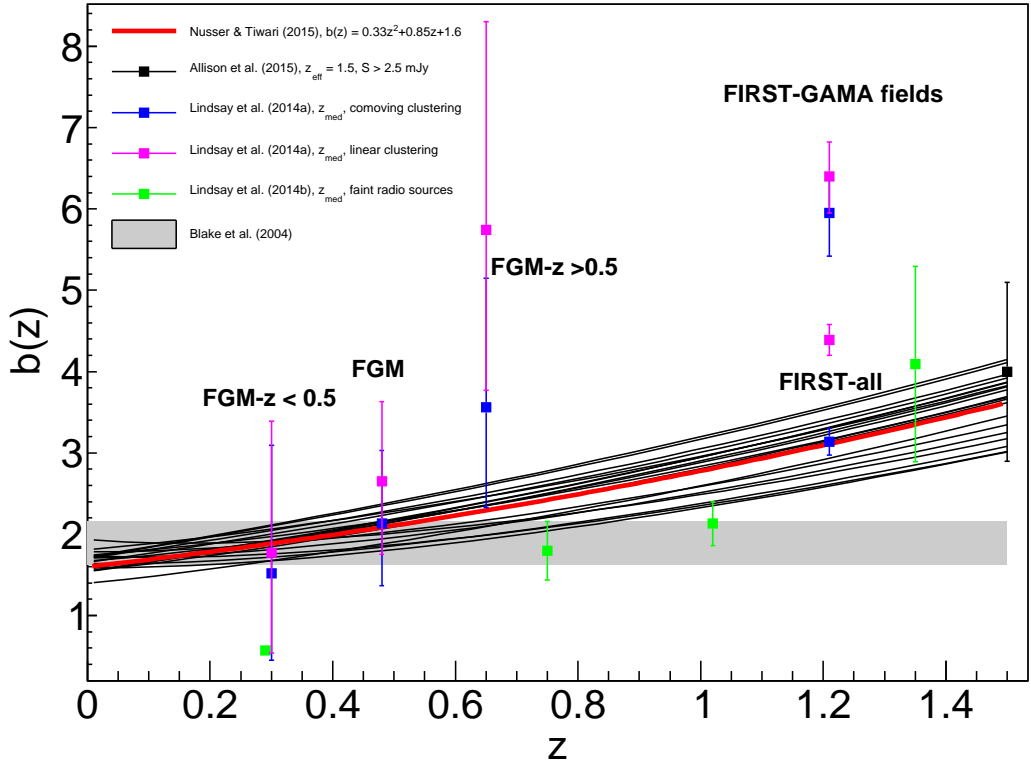}
\caption{The redshift dependent $\bnt(z)$ in equation (\ref{eq:biasingNT}) is plotted as the thick red curve for the best fit parameters. 
Thin curves correspond to random choices of the parameters appearing in $\bnt(z)$.   
For comparison, results from the literature are also shown, as indicated in the figure. 
The abbreviation FGM stands for `FIRST-GAMA-matched'.}
 \label{fig:bias}
\end{figure}

To model the biasing of radio galaxies, NT15 use an elaborate model based on halo biasing \citep{Sheth:2001} 
and the relationship between galaxy stellar masses and the prevalence of radio activities \citep{Moster:2013}. 
They provide the following convenient parametric fit. 
\begin{equation}
\label{eq:biasingNT}
\bnt(z)=b_1 z^2+b_2 z +b_3\; ,
\end{equation} 
where  the best fit values of $b_i$ are $b_1=0.33$, $b_2=0.85$ and $b_3=1.6$.
Furthermore, we shall employ  other choices for the bias factor  to fit with NVSS dipole observation.  
Particularly, we consider constant  bias factors $b(z)=2$ and $3$.
The thick  red line in figure \ref{fig:bias} is $\bnt(z)$ for the best fit parameters $b_i$, 
while the thin black lines represent 
random $b_i$ drawn from the full  probability function   from  NT15. 
Several estimates of the bias factor available in the literature are also plotted in the figure.
 The values  from Allison et al. \citep{Allison:2015} and Lindsay et al. 
\citep{Lindsay:2014a} have a large uncertainty and  agree with the bias factors  used in this work. However, 
the bias of  faint galaxies (S$_{\rm 1.4}>$ 90 $\rm \mu$Jy) \citep{Lindsay:2014b} is apparently  low.
The bias measurements in \citep{ Allison:2015,Lindsay:2014a,Lindsay:2014b} correspond to  large redshift bins and the 
redshift for each data points is the median or effective redshift of the bin. 
The bias of  NVSS galaxies  obtained by Blake et al. \citep{Blake:2004}  (grey band), 
is assumed to be constant with redshift  in their analysis and does not take into account  the dipole. 
Unfortunately, we do not have any radio galaxy bias observation at redshift $z\ltsim 0.1$ where the dipole signal is mostly formed.

\section{The Mock catalogs}
\label{sc:sim}
The origin of the intrinsic dipole signature is fluctuations on large scales with negligible  coupling to 
nonlinear evolution on  small scales. Hence it suffices for our purposes to construct mock NVSS catalogs 
based on linear theory for structure formation.  The mocks are extracted from a 
random Gaussian realization of  linear density and velocity fields sampled on a $512^3$  cubic grid  
of 17.408 comoving Gpc on the side. The  box is equivalent, in volume,  to a sphere extending to  redshift $z\sim 7$. 
The grid  spacing of 34 Mpc is well below the physical scales relevant for the dipole. 
The realization of the linear dark matter density, $\delta_m$,  in the box  is generated with 
the  {\bf GRAFIC-2} \citep{Bertschinger:2001ng} package and 
the corresponding velocity field is computed using the linear theory relation between 
velocity and density \citep{Peebles:1980}.
We adopt  the latest Planck results \citep{Planck:2015} for the cosmological parameters:
Hubble constant $H_0 = 67.8$ km s$^{-1}$ Mpc$^{-1}$, total matter density parameter $\Omega_m = 0.308$, 
baryonic density parameter $\Omega_b = 0.0486$, linear clustering amplitude on 8 $\hmpc$ scale, 
$\sigma_8 = 0.815$ and a spectral index $n_s = 0.9667$.
We use the fitting formulae from Ref.\cite{Eisenstein:1998} for the  $\Lambda$CDM 
power spectrum. ``Observers" are placed at the centers of spheres of 100 Mpc in radius and having with bulk motions 
of  $\sim300$ km s$^{-1}$. This is designed to match the observed bulk flow as measured in \citep{Adi:2011vl}. 
The mean number of the galaxies in a bin at redshift $z_{\rm bin}$, $N({\rm bin})$, is, 
\begin{equation}
N({\rm bin}) =\bar n({ z_{\rm bin}})\times dv \times [1+b(z_{\rm bin}) D(z_{\rm bin}) \delta_m] ,
\label{eq:binn}
\end{equation} 
where $\bar n(z_{\rm bin})$ is the mean number density at the bin and $dv$ is the bin volume.
The number density $\bar n(z_{\rm bin})$ is obtained from the redshift distribution
form the equation (\ref{eq:Nmodel}). The actual number of galaxies in the bin
is a random integer, drawn from a Poisson distribution with mean $N({\rm bin})$. 

\begin{figure}[!h]
\includegraphics[width=1.0\textwidth]{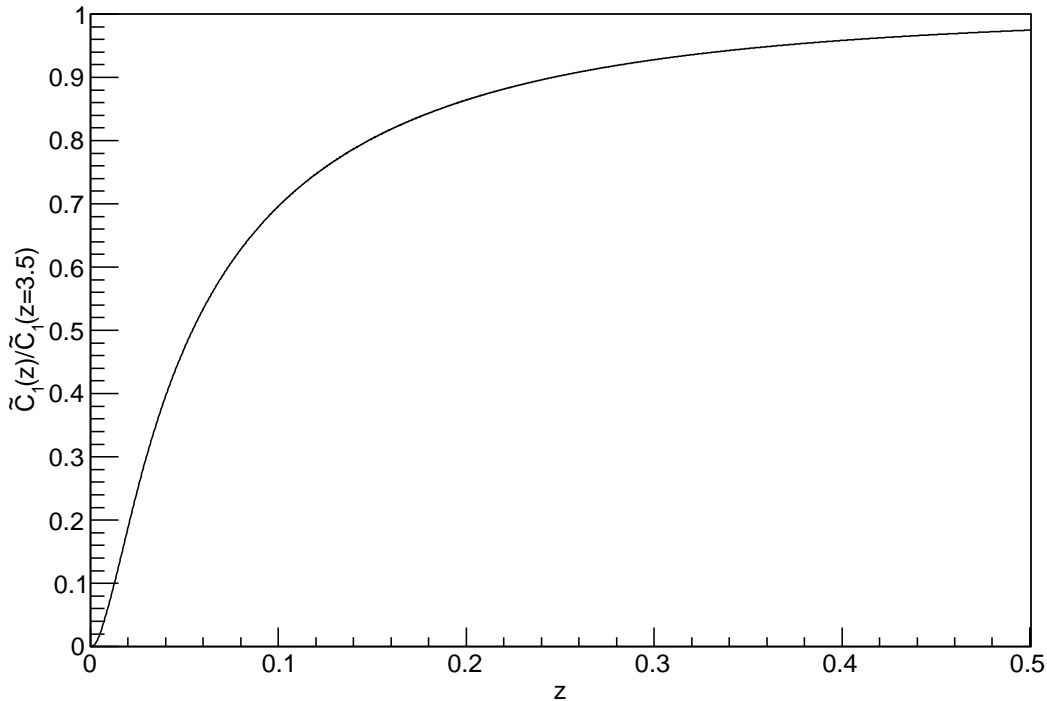}%
\caption{The fractional contribution to $\tilde{C_1}$ from structures below redshift $z$.}
 \label{fig:c1z}
\end{figure}

In order to include scatter due to uncertainties in the fits of $N(z)$ and 
for the biasing form $\bnt(z)$  in equation (\ref{eq:biasingNT}), we draw 
random  realizations of the parameters $a_i$ and $b_i$ (see equations (\ref{eq:Nmodel})  \& (\ref{eq:biasingNT})) 
from  the full probability distribution for 
the observed $C^{\rm obs}_l$ ($l>4$)  computed in NT15 (see figures \ref{fig:NzSim} \& \ref{fig:bias}).
$10^5$ NVSS mocks are generated, respectively, from these random sets rather than the mean values. 
The  biasing form $\bnt$ is  unsuitable for high redshifts but this is insignificant since the contribution 
to the dipole from large scale structure at $z>1$ practically vanishes (see figure \ref{fig:c1z}). 
For computational purposes we simply take $b(z>1.5)=b(z=1.5)$. 
When exploring constant bias values, we still generate $10^5$ mocks for random shot noise and  $a_i$.
The angular orientation  of the observed NVSS sky  in relation to the direction of the observed dipole is matched exactly. 
We rotate the mocked map such as the  dipole direction matches with the observed direction and 
then we mask the Galactic plane and 22 bright local extended sites as identified in \cite{Blake:2002}. 
These final mock NVSS galaxy maps are  used to calculate $C_{l}$ and are compared with NVSS observed dipole. 

It is important to  explore the range of redshifts and scales of structures 
which are mostly responsible for the intrinsic  dipole signature. In figure \ref{fig:c1z}
we show the fractional contribution to  $\tilde{C_1}$ from structure up to a certain redshift from the theoretical 
relation (see equation (\ref{eq:clphi})). The dipole is mostly determined by low redshift structure with 60\%  of the power already achieved by 
$z=0.1$. 

\begin{figure}
\includegraphics[width=1.0\textwidth]{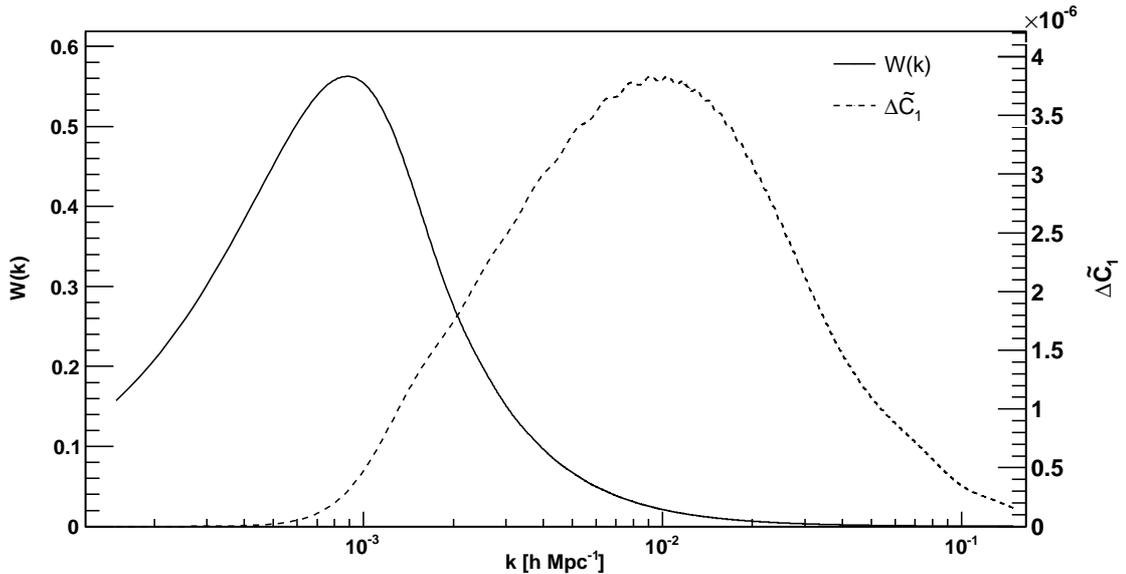}%
\caption{ The dipole $k$-space window function, $W(k)$,  and the theoretical $\tilde{C_{1}}$ par $log(k)$ bins.  
The $\Delta\tilde{C_{1}}$ integral over per $log(k)$ is normalised to $\Lambda$CDM $\tilde{C_{1}}$.
The dipole ($\tilde{C_{1}}$) is most sensitive to $k\sim 0.01$ $\hmmpc$.
}
\label{fig:window}
\end{figure}

In figure \ref{fig:window} we show the 
$k$-space window function, $W(k)$, and the theoretical differential contribution $\tilde{C_{1}}$ to per $\log(k)$ bins. 
Although the window function  peaks around $k\sim 0.001 \hmmpc$, the 
overall dipole signal gets  most of its  contribution from scales $k\sim 0.01$ $\hmmpc$. 

\section{Results} 
\label{sc:res}
The statistical assessment of the ability of the models to match  
the observed $C^{\rm obs}_1$ is done on the basis of  counting the number of mocks 
with $C_1$ exceeding the observed value. 
We perform this analysis for several lower cuts, $S_{\rm min}$,  on the flux density, $S$, where for each cut 
we use  $ 10^5$ random mocks. 
  The  observed $C^{\rm obs}_1$ is calculated after removing the contribution 
of the solar motion  from the NVSS number density maps, according to the relation (\ref{eq:D_n}).

\begin{figure}
\includegraphics[width=1.0\textwidth]{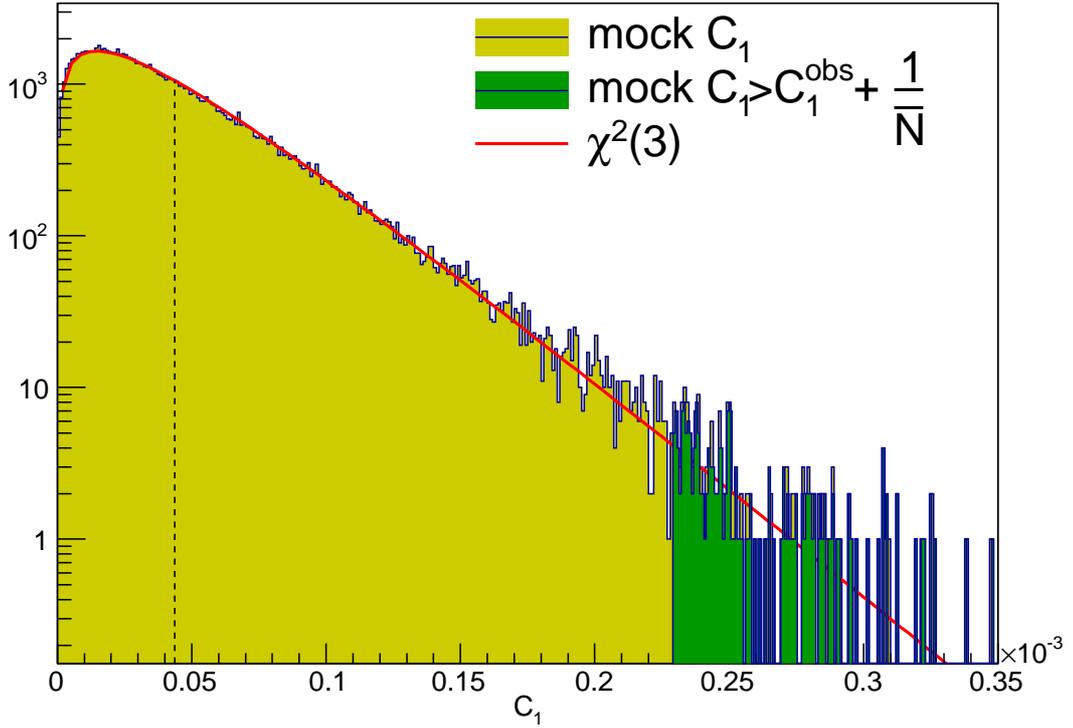}%
\caption{The mocks $C_{1}$ distribution and the observation probability from 
1,00,000 samples is shown for the flux density cut $S>20$ mJy. The dotted line is mean 
$C_{1}$. Note that the y-axis is in log scale as the probability for NVSS observed dipole 
or a dipole with larger amplitude is extremely tiny (0.2\%)
and not even visible in linear scale. A good fit to mocks $C_{1}$ distribution 
to $\chi^{2}(3)$ probability distribution is also shown.}
\label{fig:c1prob}
\end{figure}

We first explore consequences of  $b(z)=\bnt(z)$ in  (\ref{eq:biasingNT}) with the parameters which yield a good match to $C_l$ for
$l>4$ (see NT15). 
Figure \ref{fig:c1prob} shows the  distribution of $C_1$ in the $10^5$ mocks for $S_{\rm min}=20$ mJy. The distribution is well 
approximated by  a $\chi^2$ distribution with 
3 degrees of freedom corresponding to  the average of common variance of three independent Gaussian 
variables  ($<(|a_{11}|^2+ |a_{10}|^2+ |a_{11}|^2>$). 
The  theoretical   $\chi^2$ distribution is also plotted in the figure \ref{fig:c1prob}, as the red curve.  
{The  probability for the mocks to yield a $C_1$ larger than the observed NVSS dipole $C^{\rm obs}_{1}$ is 0.2\% 
for  $\bnt$.} 
This  corresponds to $\sim 2.89 \sigma$ significance level for a normal distribution of one variable.
Values of  $C^{\rm obs}_{1}$ for several flux thresholds, $S_{\rm min}$, 
are summarised in table \ref{tab:res1}. Note that the NVSS dipole  for $S_{\rm min}=10$ mJy may suffer from 
systematic biases \citep{Blake:2002} and we consider it unreliable. We list results with this cut 
only for the sake of  completeness and comparison with  other authors.
According to the results in table \ref{tab:res1}, the $\Lambda$CDM mocks with the bias $\bnt$ are
reasonably consistent at the $2.12 \sigma$ level with the NVSS dipole with $S_{\rm min}= 15 \; \rm mJy$. 
Nonetheless, there is a   tension between  the mocks and the data at higher fluxes.  
The biasing template   $\bnt$ is inferred from   high order multipoles which 
are sensitive to the galaxy  distribution at redshifts $z\gtsim 0.2$ (see figure 7 in NT15). 
The dipole signal, however, is { largely} generated at  lower redshifts, $z\lesssim0.1$  
by fluctuations with  wave numbers $k\sim 0.01 \rm h Mpc^{-1}$ (cf.  
figures  \ref{fig:c1z} \& \ref{fig:window}).
Therefore, the choice   $b(z)=\bnt$  may not be suitable for the dipole in the case of a redshift and scale dependent (radio)  galaxy biasing.
Alternatively, we generate mocks for  $b(z)=2$ and $3$.
The  results for multipoles $l=1,2,3$ and 4 are shown in table \ref{tab:resC1234}.
Both values of constant $b$ are consistent with the dipole for $S_{\rm min}=15$ 
and, as with $\bnt$, higher flux thresholds are in tension with the observed dipole for $b=2$. 
The high value $b=3$ tones down the dipole tension to an acceptable level. However, this value   
is significantly high compared to most observational estimates \citep{Allison:2015,Lindsay:2014a,Lindsay:2014b,Blake:2004,Adi:2015nb} as seen in figure \ref{fig:bias}.

\begin{table*}[!h]
   \centering
  \begin{tabular}{|c|c|c|c|c|c|c| }
    \hline 
    & & & & & & \\
$S_{\rm min} $ (mJy)  & (RA, Dec) & ($l, b$) & $C^{\rm obs}_1$ $(\times 10^4)$  &  shot-noise ($\times 10^5$) & p-value  &    $\sigma$- significance \\
    & & & & & & \\
\hline
    & & & & & & \\
 10 & (135$^{\circ}$, 23$^{\circ}$)   & (204$^{\circ}$, 38$^{\circ}$) & 0.82  & 1.8  &  0.02389  &   1.98 \\
 \hline
    & & & & & & \\
 15 &(151$^{\circ}$,  -6$^{\circ}$) &  (246$^{\circ}$, 38$^{\circ}$) & 1.10  & 2.5  &  0.01696  &   2.12 \\
 20 &(151$^{\circ}$, -14$^{\circ}$) &  (253$^{\circ}$, 32$^{\circ}$) & 1.96  & 3.3  &  0.00193  &   2.89 \\
 30 &(160$^{\circ}$, -16$^{\circ}$) &  (263$^{\circ}$, 36$^{\circ}$) & 2.37  & 4.7  &  0.00262  &   2.79 \\
 40 &(150$^{\circ}$, -35$^{\circ}$) &  (268$^{\circ}$, 16$^{\circ}$) & 2.89  & 6.3  &  0.00173  &   2.92 \\
 50 &(174$^{\circ}$, -36$^{\circ}$) &  (286$^{\circ}$, 24$^{\circ}$) & 3.79  & 8.0  &  0.00209  &   2.86 \\
 \hline          
\end{tabular}
\caption{The observed $C^{\rm obs}_1$ and shot-noise ($\frac{1}{\bar \cN}$) 
along with dipole direction. The probability is the count of mocks with $C_1$ exceeding  
the observed  $C^{\rm obs}_1$. The corresponding $\sigma$- significance is also given for the observed 
probability.
The p-value is the probability ($C_1>C^{\rm obs}_1$).}
\label{tab:res1}
\end{table*}

\begin{table*}
\footnotesize
   \centering
  \begin{tabular}{|c|c|c|c|c|c|c|c|c| }
    \hline
  &\multicolumn{2}{c|}{} &\multicolumn{2}{c|}{} &\multicolumn{2}{c|}{} &\multicolumn{2}{c|}{}\\
$S_{\rm min} $ (mJy) & \multicolumn{2}{c|}{$C_{1}$} &\multicolumn{2}{c|} {$C_{2}$} &\multicolumn{2}{c|} {$C_{3}$} &\multicolumn{2}{c|} {$ C_{4}$} \\
\hline
    &p-value&$\sigma$-significance&p-value&$\sigma$-significance &p-value&$\sigma$-significance &p-value&$\sigma$-significance \\
\hline
\multicolumn{9}{|c|}{}\\
\multicolumn{9}{|c|}{ $b(z)=2.0$}\\
\multicolumn{9}{|c|}{}\\
 15 &0.01437 &2.19 &0.00324 & 2.72 &0.32412 & 0.46 & 0.12362 &1.16 \\
 20 &0.00149 &2.97 &0.00493 & 2.58 &0.96881 & ---  & 0.03533 &1.81 \\
 30 &0.00300 &2.75 &0.00765 & 2.43 &0.69182 & ---  & 0.30641 &0.51 \\
 40 &0.00252 &2.80 &0.00696 & 2.46 &0.41254 & 0.22 & 0.41076 &0.23 \\
 50 &0.00154 &2.96 &0.08581 & 1.37 &0.46751 & 0.08 & 0.31316 &0.49 \\
\multicolumn{9}{|c|}{}\\
\multicolumn{9}{|c|}{$b(z)=3.0$}\\
\multicolumn{9}{|c|}{}\\
 15 &0.06170 &1.54 &0.02345 &1.99 &0.56519 & ---  &0.34328 & 0.40 \\
 20 &0.00922 &2.36 &0.02290 &2.00 &0.98633 & ---  &0.13678 & 1.10 \\
 30 &0.01011 &2.32 &0.02437 &1.97 &0.80103 & ---  &0.49466 & 0.01 \\
 40 &0.00656 &2.48 &0.01882 &2.08 &0.53594 & ---  &0.56074 & --- \\
 50 &0.00391 &2.66 &0.13536 &1.10 &0.57366 & ---  &0.43821 & 0.16 \\
 \hline
\end{tabular}
\caption{The low multipole $C_1$, $C_2$, $C_3$ and $C_4$ for different 
biasing templates. The p-value is the probability ($C_l>C^{\rm obs}_l$). For all 
p-values ($<0.5$) we have given the corresponding $\sigma$-significance for a normal distribution 
of one variable.}
\label{tab:resC1234}
\end{table*}
\normalsize

\section{Conclusion and Discussion } 
\label{sc:con}

We have presented an analysis of  the NVSS dipole incorporating, for the first time, the 
contribution from the large scale inhomogeneities in the galaxy distribution. 
Mock catalogs mimicking the NVSS catalog are used to model uncertainties in the galaxy  redshift distribution 
and incomplete angular coverage of the data. 
The bias factor obtained from $l\geq4$ by NT15  is at odds with the number count dipole measurement  at the  $\sim2.9 \sigma$ 
significance level for the $\Lambda$CDM cosmology. 
However,  the intrinsic dipole signal is shapes at low ($z\ltsim 0.1$) redshifts and 
on large scales ($k\sim 0.01 \rm h Mpc^{-1}$); about $70\%$ ($45\%$) of the dipole ($l=4$) contribution is from 
redshift $z\lesssim0.1$. Therefore, it is prudent to consider a large   bias factor for radio galaxies at 
low redshifts  as an explanation to the  dipole signal.
Independent  bias estimates  at such low redshift are crude, as the redshift of radio galaxies
is generally unknown and we do not have much galaxies in low redshift bins.
A direct assessment of the redshift dependence  of the bias of radio galaxies is done 
in wide redshift bins due to the small number of radio galaxies with redshift information.
A summary of some bias factor determination in the literature is given figure \ref{fig:bias}.  
The radio galaxy bias at low redshift $z\lesssim0.1$ is loosely constrained and 
if the bias is as high as $b(z\lesssim 0.1)\approx 3 $, it  tones down the disagreement between the 
model and the data to a $\sim 2.3 \sigma$ significance level  for $S>20$, 30, 40 and 50 mJy.
This  bias value   agrees, within the large error bars, with the estimate 
obtained in Ref.\cite{Lindsay:2014a} (see figure \ref{fig:bias}).  But  it is substantially  
high  with respect to other results  \citep{ Allison:2015,Lindsay:2014b,Blake:2004}, including 
the value obtained by NT15  for the same NVSS data from higher multipoles $l\ge4$. We emphasize that the 
bias extracted in NT15 follows the detailed halo biasing recipe  \cite{Sheth:2001} and 
a relation between galaxy stellar mass and radio activity \citep{Moster:2013}.

The NVSS galaxy sky  distribution may be affected by the 
presence of a large scale mass distribution which generates 
the  observed bulk motion of  a sphere of radius $100\hmpc$ around the observer. 
The modeling here incorporates the presence of this mass fluctuation by selecting ``observers" satisfying the bulk motion 
constraint.  We also rotate the maps so that the dipole signal is aligned with the observation mask as in the real data. 
As seen in  table \ref{tab:compvr},  these two constraints give modest improvement in matching the model to the 
observations. 

There are certain claims in the literature for excess of power on  scales of 100s of Mpcs relative to the
$\Lambda$CDM model \cite{Watkins:2009,Macaulay:2011,Thomas:2011}. We will refrain from a detailed assessment 
of these claims and we simply emphasize the following point. The dipole signal is matched 
for the $\Lambda$CDM for $b=3$. Therefore, any such power excess must be 3 fold higher than 
the $\Lambda$CDM. Such large excess at $z\sim 0.1$ will generate a bulk flow within a $100\hmpc$ which 
is 3 times the predicted value of the bulk flow in the $\Lambda$CDM, i.e. a motion of $\sim 800 \kms$. 
Such a value of the bulk flow is at odds with the highest bulk flow estimates in the literature. 

There are several non-conventional explanation for the NVSS dipole. 
 For example, the anisotropic mode generated before inflation may re-enter the horizon 
at late time and produce large scale anisotropy \citep{Aluri:2012,Rath:2013,Shamik:2013}. 
Alternatively, the primordial non-Gaussianities (NG) may affect the structures at large scales.  
Particularly, significant NG in inflationary cosmology scenario may potentially give an 
increase in biasing over large scale and produce excess dipole signal \cite{Dalal:2008}. 
In Ref. \cite{Rubart:2014} the authors argue that NVSS dipole anisotropy can be partially 
explained with a local over- or under density volume. The authors in Ref. \cite{Rubart:2014}
constrain themselves to local structures consistent with CMBR dipole and other observations.

The NVSS dipole remains puzzling within the ambit of radio galaxy biasing. 
The extra-large power required to explain dipole signal is $\sim3$ fold in comparison to the $\Lambda$CDM and 
very unlikely. The NVSS excess dipole is most likely  due to a residual calibration error across the sky 
which also affects the high flux cuts.
The radio galaxy dipole observation will immensely improve with the Square Kilometer Array.
\begin{table*}
   \centering
  \begin{tabular}{|c|c|c|c| }
    \hline
    & & &  \\
 $S_{\rm min} $ (mJy) & 15 & 20 & 30  \\
    \hline
\multicolumn{4}{|c|}{}\\
\multicolumn{4}{|c|}{considering all constraints}\\
\multicolumn{4}{|c|}{}\\
 p-value & 0.01696 &  0.00193 & 0.00262  \\
 $\sigma$- significance &  2.12  & 2.88 & 2.79  \\
\multicolumn{4}{|c|}{}\\
\multicolumn{4}{|c|}{No velocity constraint}\\
\multicolumn{4}{|c|}{}\\
 p-value & 0.01625 & 0.00175 & 0.00224  \\
 $\sigma$- significance &  2.14  & 2.92 & 2.84  \\
\multicolumn{4}{|c|}{}\\
\multicolumn{4}{|c|}{No rotation to map}\\
\multicolumn{4}{|c|}{}\\
 p-value & 0.01280 & 0.00136 & 0.00238  \\
 $\sigma$- significance &  2.23  & 3.00 & 2.82  \\
 
\hline
\end{tabular}
\caption{The dependence of $C_1$ to different constraints used in mocks. 
The results are show for flux density cut $S>$ 15, 20 and 30 mJy.}
\label{tab:compvr}
\end{table*}

\section{Acknowledgments}
We thank Ranieri Baldi for discussion and Curtis Saxton for reading the manuscript.
This research was supported by the I-CORE Program of the Planning and Budgeting
Committee, THE ISRAEL SCIENCE FOUNDATION (grants No. 1829/12 and No. 203/09), 
the Asher Space Research Institute, and the Munich Institute for Astro and Particle Physics (MIAPP) 
of the DFG cluster of excellence Origin and Structure of the Universe.
This work is also supported in part at the Technion by a fellowship from the Lady Davis Foundation.
We have used CERN ROOT 5.34/21 \citep{root} for generating our plots. 
\bibliographystyle{JHEP}
\bibliography{master,NVSS}
\end{document}